\documentclass[english,journal=jpcafh,manuscript=article,layout=twocolumn]{achemso}
\usepackage[T1]{fontenc}
\usepackage[utf8]{inputenc}
\usepackage{babel}
\usepackage{url}
\usepackage{amsmath}
\usepackage{graphicx}
\usepackage[numbers]{natbib}
\usepackage[pdfusetitle,
 bookmarks=true,bookmarksnumbered=false,bookmarksopen=false,
 breaklinks=false,pdfborder={0 0 1},backref=false,colorlinks=false]
 {hyperref}

\makeatletter

\title{OpenOrbitalOptimizer---a reusable open source library for self-consistent
field calculations}

\author{Susi Lehtola}

\email{susi.lehtola@alumni.helsinki.fi}

\affiliation{Department of Chemistry, University of Helsinki, P.O. Box 55, FI-00014
University of Helsinki, Finland}

\author{Lori A. Burns}

\affiliation{Center for Computational Molecular Science and Technology, School
of Chemistry and Biochemistry, Georgia Institute of Technology, Atlanta,
Georgia 30332, USA}

\providecommand{\tabularnewline}{\\}

\usepackage[version=4]{mhchem}
\usepackage{notes2bib}
\usepackage{upgreek}

\usepackage{cleveref}
\AtBeginDocument{%
\let\ref\cref
}

\SectionNumbersOn
\setkeys{acs}{articletitle=true}
\setkeys{acs}{doi = false} 
\setkeys{acs}{maxauthors=10}
\setkeys{acs}{etalmode=truncate}

\makeatother

\begin{document}
\begin{abstract}
According to the modern paradigms of software engineering, standard
tasks are best accomplished by reusable open source libraries. We
describe OpenOrbitalOptimizer: a reusable open source C++ library
for the iterative solution of coupled self-consistent field (SCF)
equations $\boldsymbol{F}_{p}^{\sigma}(\{\boldsymbol{C}_{p}^{\sigma}\})\boldsymbol{C}_{p}^{\sigma}=\boldsymbol{C}_{p}^{\sigma}\boldsymbol{E}_{p}^{\sigma}$
for an arbitrary number of particle types $p$ and symmetries. Although
OpenOrbitalOptimizer is a new project, it already implements standard
algorithms for solving SCF equations: Pulay's direct inversion in
the iterative subspace (DIIS), energy DIIS (EDIIS), augmented DIIS
(ADIIS), and the optimal damping algorithm (ODA). The library was
designed as an easy way to introduce the state-of-the-art convergence
accelerators in a number of legacy programs. It is easy to interface
with various programs, as it only requires a function to evaluate
the total energy $E$ and Fock matrices $\{\boldsymbol{F}_{p}^{\sigma}\}$
for a given set of orbitals $\{\boldsymbol{C}_{p}^{\sigma}\}$. The
only assumption behind the library is that one is able to easily store
Fock and orbital matrices in memory, and to diagonalize the Fock matrices
in full, which is the case in the overwhelming majority of quantum
chemistry applications. We exemplify the library with nuclear-electronic
orbital (NEO) calculations of protonated water clusters with Gaussian-type
orbital basis sets. We find that a minimal-basis protonic guess works
well, and that the stepwise SCF algorithm requires less computational
time than the simultaneous SCF algorithm.
\end{abstract}
\newcommand*\ie{{\em i.e.}}
\newcommand*\eg{{\em e.g.}}
\newcommand*\etal{{\em et al.}}
\newcommand*\citeref[1]{ref. \citenum{#1}}
\newcommand*\citerefs[1]{refs. \citenum{#1}} 

\newcommand*\Erkale{{\sc Erkale}}
\newcommand*\Bagel{{\sc Bagel}}
\newcommand*\FHIaims{{\sc FHI-aims}}
\newcommand*\LibXC{{\sc LibXC}}
\newcommand*\Orca{{\sc Orca}}
\newcommand*\PySCF{{\sc PySCF}}
\newcommand*\PsiFour{{\sc Psi4}}
\newcommand*\Turbomole{{\sc Turbomole}}

\section{Introduction \label{sec:Introduction}}

Electronic structure calculations have become widely used in academia
and industry to explain and predict the properties of materials, and
the various steps involved in chemical reactions, for example. A typical
electronic structure calculation involves finding the electronic one-particle
states also known as orbitals. These orbitals are usually found by
minimizing the Hartree--Fock or Kohn--Sham density functional\citep{Hohenberg1964_PR_864,Kohn1965_PR_1133}
for the total energy $E[\left\{ \psi_{i\sigma}\right\} ]$,\citep{Lehtola2020_M_1218}
which depends on the occupied set of the orbitals $\psi_{i\sigma}$
of spin $\sigma$. Depending on the type of the system, the orbitals
$\psi_{i\sigma}$ are called atomic, molecular, or crystalline orbitals
for atoms, molecules, and periodic systems, respectively. 

To allow for a computational solution, the orbitals must be discretized
in some form. The typical choice in quantum chemistry is the linear
combination of atomic orbitals (LCAO) method, especially when combined
with the use of Gaussian-type orbital (GTO) basis functions.\citep{Lehtola2019_IJQC_25968}
In LCAO, the orbitals are written in the basis set $\chi_{\mu}(\boldsymbol{r})$
as
\begin{equation}
\psi_{i\sigma}(\boldsymbol{r})=\sum_{\alpha}C_{\mu i}^{\sigma}\chi_{\mu}(\boldsymbol{r}),\label{eq:basexp}
\end{equation}
and the energy is now minimized with respect to the orbital expansion
coefficients $\boldsymbol{C}^{\sigma}$. However, since \ref{eq:basexp}
obviously contains no assumption on the form of the basis functions,
it can also be used in combination with other types of basis sets,
such as finite elements, for example.

The variation of the total energy functional with respect to the orbital
coefficients leads to the Roothaan-type equation\citep{Roothaan1951_RMP_69,Pople1954_JCP_571,Berthier1954_JCP_363,Lehtola2020_M_1218}
\begin{equation}
\boldsymbol{F}^{\sigma}(\{\boldsymbol{P}^{\sigma}\})\boldsymbol{C}^{\sigma}=\boldsymbol{S}\boldsymbol{C}^{\sigma}\boldsymbol{E}^{\sigma},\label{eq:roothaan-nonorth}
\end{equation}
where $F_{\mu\nu}^{\sigma}=\partial E/\partial P_{\mu\nu}^{\sigma}$
is the Fock matrix, $S_{\mu\nu}=\langle\mu|\nu\rangle$ is the overlap
matrix of the basis set, $\boldsymbol{E}^{\sigma}$ is a diagonal
matrix holding the orbital energies, and 
\begin{equation}
P_{\mu\nu}^{\sigma}=\sum_{i}f_{i\sigma}C_{\mu i}^{\sigma}(C_{\nu i}^{\sigma})^{\dagger}\label{eq:densmat}
\end{equation}
is the density matrix where $f_{i\sigma}$ are the orbital occupation
numbers. Any solution to \ref{eq:roothaan-nonorth} is a critical
point of the total energy; typically, one hopes that the solution
is a local (if not the global) minimum, but the self-consistent solution
of \ref{eq:roothaan-nonorth} can also converge to a saddle point.

The central task now becomes to find such a set of orbitals $\boldsymbol{C}^{\sigma}$
that generates Fock matrices $\boldsymbol{F}^{\sigma}$ that yield
the same $\boldsymbol{C}^{\sigma}$ upon the solution of \ref{eq:roothaan-nonorth};
this is known as the self-consistent field (SCF) problem.\citep{Lehtola2020_M_1218}
The SCF method consists of an iterative procedure: starting from some
initial guess for the orbitals $\boldsymbol{C}^{\sigma}$ (which can
be also obtained from a guess for the Fock matrices $\boldsymbol{F}^{\sigma}$,
or density matrices $\boldsymbol{P}^{\sigma}$ that will be used to
build such guess Fock matrices),\citep{Almloef1982_JCC_385,VanLenthe2006_JCC_32,Lehtola2019_JCTC_1593,Lehtola2020_JCP_144105}
one computes a new Fock matrix, which in turn determines a new set
of orbitals.

The Roothaan equation is typically solved by re-expressing the unknown
orbitals in an orthonormal basis set\citep{Lehtola2020_M_1218}
\begin{equation}
\boldsymbol{C}^{\sigma}=\boldsymbol{X}\boldsymbol{\tilde{C}}^{\sigma}\label{eq:Corth}
\end{equation}
and left-projecting \ref{eq:roothaan-nonorth} with $\boldsymbol{X}^{\text{T}}$,
which reduces the problem to a normal eigenvalue equation
\begin{equation}
\boldsymbol{\tilde{F}}^{\sigma}(\{\boldsymbol{P}^{\sigma}\})\tilde{\boldsymbol{C}}^{\sigma}=\tilde{\boldsymbol{C}}^{\sigma}\boldsymbol{E}^{\sigma}\label{eq:roothaan}
\end{equation}
where 
\begin{equation}
\tilde{\boldsymbol{F}}=\boldsymbol{X}^{\text{T}}\boldsymbol{F}\boldsymbol{X}\label{eq:Forth}
\end{equation}
in the case that $\boldsymbol{X}$ is chosen such that $\boldsymbol{X}^{\text{T}}\boldsymbol{S}\boldsymbol{X}=\boldsymbol{1}$.
Such matrices $\boldsymbol{X}$ can be formed with symmetric orthogonalization,\citep{Loewdin1950_JCP_365}
or canonical orthogonalization when the basis set has linear dependencies.\citep{Loewdin1956_AP_1}
If the basis set contains many linear dependencies, a smaller basis
set that spans the same space can be first found with a pivoted Cholesky
decomposition before carrying out the conventional orthonormalization
procedure.\citep{Lehtola2019_JCP_241102,Lehtola2020_PRA_32504}

\Crefrange{eq:Corth}{eq:Forth} define the Roothaan update, enabling
one to determine new orbitals from the obtained Fock matrix. However,
the central issue in SCF calculations is that this procedure of iterative
Fock builds and diagonalizations is insufficient for all but the simplest
of systems. 

To begin, the Roothaan procedure for Hartree--Fock can be shown to
either converge to a critical point of the Hartree--Fock functional
(\emph{i.e.}, a solution of \ref{eq:roothaan-nonorth}), or to converge
asymptotically to a pair of density matrices, each iteration alternating
between the two in an effect commonly known as \emph{charge sloshing}
(see \citeauthor{Cances2000_EMMNA_749}\citep{Cances2000_EMMNA_749,Cances2000__17}
for the proof). For this reason, SCF convergence has typically to
be stabilized in the early iterations in order to damp down this sloshing. 

Another issue is that even when in case where the Roothaan iterations
converge onto a critical point of the functional, the convergence
can be too slow for practical applications. Methods to extrapolate
the Roothaan updates are therefore desirable.

Various schemes have been developed to address these two issues in
SCF calculations. As the literature is very broad, we will only present
the most important aspects of the literature here, postponing the
more thorough discussion to a dedicated review article.

While convergence acceleration near the converged solution has long
been typically handled with the direct inversion in the iterative
subspace\citep{Pulay1980_CPL_393} (DIIS) approach of \citet{Pulay1982_JCC_556},
determining a wave function that is sufficiently close to the solution
for DIIS to be applicable is traditionally the harder problem, especially
when accurate initial guesses\citep{Almloef1982_JCC_385,VanLenthe2006_JCC_32,Lehtola2019_JCTC_1593,Lehtola2020_JCP_144105}
are not available in the employed electronic structure program: when
a poor quality guess such as the one-electron guess obtained by omitting
all electronic interactions from the Fock matrix is used, the guess
Fock matrix and the orbitals are far from their optimal values, and
significant changes in the orbitals and their occupations are typically
observed in the first iterations.

The traditional approach to stabilizing convergence is to damp the
density matrix update; an optimal damping algorithm has been suggested
by \citet{Cances2000_IJQC_82}. Level shifting\citep{Saunders1973_IJQC_699}
is another commonly used alternative, and it has been shown\citep{Cances2000_EMMNA_749,Cances2000__17}
that a sufficiently large level shift always leads to progress towards
the minimum; however, the rate of convergence also becomes the slower
the larger the shift is.

Newer approaches to allow robust SCF convergence include using physically
motivated proxy energy functionals in combination with a DIIS type
approach,\citep{Kudin2002_JCP_8255,Hu2010_JCP_54109} as well as various
procedures based on the direct minimization of the energy with respect
to orbital rotations.\citep{Chaban1997_TCA_88,HeadGordon1988_JPC_3063,Neese2000_CPL_93,VanVoorhis2002_MP_1713,VandeVondele2003_JCP_4365,Francisco2004_JCP_10863,Host2008_JCP_124106,Host2008_PCCP_5344,Levi2020_FD_448,Slattery2024_PCCP_6557,Sethio2024_JPCA_2472}

Still, a major problem in practical electronic structure calculations
today is the lack of state-of-the-art SCF algorithms in many program
packages, which poses severe issues for ease-of-use and computational
efficiency of these packages. While state-of-the-art SCF algorithms
can often quickly and reliably converge even challenging electronic
structures, such as those found in transition metal complexes, traditional
SCF algorithms may require hundreds more SCF iterations to reach convergence,
or fail to converge at all, necessitating user action.

A further complication is that many packages lack good initial guesses,
which makes the orbital optimizations much more challenging than when
a qualitatively correct initial guess is employed. Atomic guesses
are typically already close to the DIIS convergence region, but sadly,
many programs focusing on post-Hartree--Fock calculations are limited
to the guess corresponding to the one-electron (core) Hamiltonian.

The central issue behind these limitations is that initial guesses
and orbital optimizers are commonly deeply tied to the electronic
structure package. In fact, the base layer of any electronic structure
program habitually consists of the integrals and the SCF implementation.
This tight embedding arises from the traditional monolithic development
model of electronic structure software, which has many downsides,
as discussed recently by \citet{Oliveira2020_JCP_24117} and \citet{Lehtola2023_JCP_180901},
for example: it leads to an approach that is both laborious and error
prone, as new algorithms always need to be reimplemented from scratch
in every package. The considerable duplication of work makes it hard
to propagate improved algorithms to various program packages.

The lack of standard implementations also causes issues with the lack
of reproducibility between program packages. As discussed by \citet{Hinsen2014_F_101}
and \citet{Lehtola2022_WIRCMS_1610}, software implementations of
algorithms tend to be much more complicated than their journal descriptions,
containing many aspects and features which are never discussed in
the article. These aspects are brought to the open only if the implementation
is open source, in that it is freely available at no cost for whatever
use by whomever.\citep{Lehtola2022_WIRCMS_1610} As the ``war over
supercooled water''\citep{Smart2018_PT_} discussed by \citet{Lehtola2022_WIRCMS_1610}
has showed in practice, falsifiability of the results---an elementary
requirement of the scientific method---is only possible if the source
code used to obtain the results is freely and openly available. 

Orbital optimization and SCF algorithms are extremely good examples
of algorithms whose implementation is complicated enough that many
aspects of the actual implementations cannot be found in journal articles.
Although the field remains active in the literature,\citep{Dittmer2023_JCP_134104,Feldmann2023_JCTC_856,Qin2024_JCTC_8921,Slattery2024_PCCP_6557,Sethio2024_JPCA_2472,Oshima2025_JCP_14108}
it is often difficult to compare to the state of the art as many implementations
are in programs that are not openly available.

However, reusable open source software offers a solution.\citep{Lehtola2023_JCP_180901}
Following the discussion in \citeref{Lehtola2023_JCP_180901}, reusable
libraries can be seen as the pinnacle of scientific software development:
supplying a general solution to an important scientific problem (always
within some limited scope) shows that one has both identified and
solved the problem. Although the state of the art in the solution
techniques can change as new methods are introduced in the literature,
reusable open source libraries best satisfy the requirements of the
art of the trade (good scientific practices), as open source library
implementations can be kept more easily up to date than duplicate
implementations in individual program packages. The open source aspect
is important here (see \citet{Hocquet2021_EJPS_38}, \citet{Lehtola2022_WIRCMS_1610}
and \citet{Lehtola2023_JCP_180901} for discussion): only when the
implementation is freely accessible to \emph{all potential users}
and it can be studied and changed at source code level without having
to sign a non-disclosure agreement has the problem been actually solved;
there is a close analogy here to the push towards open access publishing
of scientific articles, which has already become a generally accepted
goal in many countries around the world.

In this work, we describe OpenOrbitalOptimizer:\citep{Lehtola2025__}
a reusable open source C++ library for orbital optimization and SCF
calculations. As far as we are aware, OpenOrbitalOptimizer is the
first of its kind: a flexible SCF library designed to be interoperable
with a large number of electronic structure packages, and compatible
with a wide variety of electronic structure methods. Previous efforts
at reusable libraries for SCF calculations appear to be limited to
the diagonalization step: for instance, the eigenvalue solvers for
petascale applications (ELPA) library,\citep{Marek2014_JPCM_213201}
the orbital minimization method (OMM) library,\citep{Corsetti2014_CPC_873}
as well as the electronic structure solver infrastructure\citep{Yu2018_CPC_267,Yu2020_CPC_107459}
(ELSI) all target the solution of a single step of \ref{eq:roothaan-nonorth}
(possibly combined with \ref{eq:densmat}), instead of the self-consistent
solution. 

OpenOrbitalOptimizer grew out of a realization that the current situation
is untenable: for instance, the orbital optimizers in ERKALE,\citep{Lehtola2012_JCC_1572}
HelFEM,\citep{Lehtola2019_IJQC_25944,Lehtola2019_IJQC_25945,Lehtola2020_PRA_12516}
Psi4,\citep{Smith2020_JCP_184108} PySCF,\citep{Sun2020_JCP_24109}
and OpenMolcas\citep{Manni2023_JCTC_6933} are all disparate, even
though they all use similar algorithms with similar implementations.
The implementation of novel algorithms separately in every program
package implies significant effort---and potential for bugs---and
the main goal of this work is to demonstrate that orbital optimization
\emph{can} be outsourced to a dedicated reusable library.

The specific reason to develop OpenOrbitalOptimizer now is our push
towards new quantum chemistry software employing numerical atomic
orbitals (NAO),\citep{Lehtola2019_IJQC_25968,Lehtola2019_IJQC_25945,Lehtola2020_PRA_12516,Lehtola2023_JCTC_2502,Lehtola2023_JPCA_4180,Aastroem2025_JPCA_2791}
which will be based on reusable library implementations to the maximal
degree possible: rather than developing a traditional monolithic program,
we realized that a focus on reusable components can have a much more
wide ranging impact,\citep{Lehtola2023_JCP_180901} since the new
reusable components can be useful also in existing program packages.

A reusable SCF implementation in a modular open source library allows
the use of state-of-the-art orbital optimizers across various electronic
structure packages. Importantly, the orbital optimization algorithm
itself is typically not a bottleneck of the calculation; instead,
most of the time in calculations is usually spent on evaluating the
total energy and the Fock matrix, \emph{i.e.}, orbital gradient for
a given set of orbitals. A smarter choice for the orbital optimizer
can significantly reduce the number of steps needed to converge onto
a (local) minimum, thus saving (potentially a considerable amount
of) computing resoures.

In addition, a reusable SCF implementation can also be easily interfaced
to legacy programs. Even though modifications to the deep base layers
of the existing SCF frameworks of legacy programs tend to be unattractive,
a reusable SCF implementation with state-of-the-art solvers can still
be used to predetermine converged orbitals, which can then be fed
into the pre-existing SCF code as an initial guess. The legacy SCF
implementation will simply reach convergence with a single step, bypassing
the need to make laborious modifications to or refactors of the legacy
code.

Furthermore, a smarter SCF implementation can make calculations considerably
easier to run, which is especially important in the current age of
big data where artificial intelligence applications may require running
millions of SCF calculations to supply training data to machine learning
models. Abstraction of SCF algorithms across various electronic structure
programs can eliminate the need to set up convoluted workflows in
case a given SCF algorithm fails to work for some data points. The
traditional solutions of changing the initial guess, adjusting the
density damping factor or level shift \emph{etc.} simply do not scale
to massive data applications.

Finally, a reusable SCF implementation can also greatly simplify the
introduction of improved initial guesses in both novel and legacy
programs. The implementation of the atomic density guess employing
fractional occupations, for example, requires extensive changes to
the SCF logic;\citep{Lehtola2020_PRA_12516} however, the guess is
easy to implement if a flexible SCF implementation is available.

A key motivation for the development of OpenOrbitalOptimizer are the
recent developments in quantum chemistry for calculations in strong
magnetic fields, where complex-valued orbital solvers become necessary
in general,\citep{Lehtola2020_MP_1597989,Aastroem2023_JPCA_10872}
as well as multicomponent calculations within the nuclear-electronic
orbital (NEO) method\citep{Webb2002_JCP_4106,Pavosevic2020_CR_4222,HammesSchiffer2021_JCP_30901}
that require the simultaneous solution of Schrödinger equations for
electrons and protons, or other quantum particles. Although the equations
can be solved by alternating between the solution of the electronic
and protonic equations, since the problems are strongly coupled, it
has been shown that the simultaneous solution converges faster.\citep{Liu2022_JPCA_7033,Feldmann2023_JCTC_856}
A key design criterion of OpenOrbitalOptimizer was thus to also enable
facile implementation of complex-valued SCF procedures as well as
NEO-type models, which feature other types of quantum particles in
addition to electrons.

The organization of the manuscript is the following. Next, we will
present the theory of SCF calculations in \ref{sec:Theory}. Then,
we will discuss the basic design choices made in OpenOrbitalOptimizer
in \ref{sec:Design-Choices}, and present the computational details
of the calculations performed in this work in \ref{sec:Computational-Details}.
We will discuss the results demonstrating the novel implementation
in \ref{sec:Results}, and finish with a summary and conclusions in
\ref{sec:Summary-and-Conclusions}.

\section{Methods \label{sec:Theory}}

We start out by simplifying the notation. As even in the case of a
non-orthonormal basis the solution of \ref{eq:roothaan} is carried
out as a standard eigenvalue problem through \cref{eq:Corth,eq:Forth},
we write \ref{eq:roothaan} in the form 
\begin{equation}
\boldsymbol{F}^{\sigma}\left(\{\boldsymbol{C}^{\sigma}\}\right)\boldsymbol{C}^{\sigma}=\boldsymbol{E}^{\sigma}\boldsymbol{C}^{\sigma}\label{eq:std-eig-notilde}
\end{equation}
where it is now implied that the quantities are defined in the orthonormal
basis. We underline that this is not a restriction, since in practice
SCF convergence acceleration has to be carried out in the orthonormal
basis, anyway.\citep{Pulay1980_CPL_393,Pulay1982_JCC_556} OpenOrbitalOptimizer
works exclusively in a basis of orthonormal orbitals, which has the
benefit of making the code simpler. 

Next, we can make a number of observations that allow us to generalize
the implementation as well as make it more powerful and efficient.
First, we realize that the spin dependence in the Roothaan equations
can be generalized to the case of an arbitrary number of types of
particles $p$
\begin{equation}
\boldsymbol{F}_{p}\left(\{\boldsymbol{C}_{p}\}\right)\boldsymbol{C}_{p}=\boldsymbol{E}_{p}\boldsymbol{C}_{p}.\label{eq:particle-roothaan}
\end{equation}
The spin-restricted case features a single type of particle (electrons),
whereas spin-polarized calculations feature two types of particles:
one for the spin-up electrons, and another for the spin-down electrons.
The huge benefit of this generalization is that also multicomponent
methods such as NEO also fit seamlessly in the library. Spin-restricted
NEO has two types of particles: the spin-restricted electrons, and
the quantum protons, which are usually assumed to be in a high-spin
state.\citep{Webb2002_JCP_4106,Pavosevic2020_CR_4222,HammesSchiffer2021_JCP_30901}
Spin-unrestricted NEO again splits the electrons into a further two
types (spin-up and spin-down electrons), so there are three types
of particles in total. The implementation supports an arbitrary number
of types of particles, and a single loop over the particle types suffices
to solve all the Roothaan equations.

Second, if symmetry is employed, the Fock matrices have a block-diagonal
structure
\[
\boldsymbol{F}=\left(\begin{array}{ccc}
\boldsymbol{F}_{\text{symm 1}} & \boldsymbol{0} & \boldsymbol{0}\\
\boldsymbol{0} & \ddots & \boldsymbol{0}\\
\boldsymbol{0} & \boldsymbol{0} & \boldsymbol{F}_{\text{symm }n}
\end{array}\right),
\]
the off-diagonal blocks vanishing as they would mix symmetries. This
means that the solution of the Roothaan equations for each type of
particle can be split into its symmetry blocks, each of which yield
corresponding orbital coefficients and orbital energies. The most
typical symmetry employed in quantum chemistry packages is Abelian
symmetry, which results in the maximum of 8 symmetry blocks corresponding
to the $A_{g}$, $B_{1g}$, $B_{2g}$, $B_{3g}$, $A_{u}$, $B_{1u}$,
$B_{2u}$, and \textbf{$B_{3u}$} blocks in the highest $D_{2h}$
symmetry; Abelian symmetries can also be found at finite magnetic
fields.\citep{Pausch2021_JCP_201101} Other examples include atomic
symmetry ($s$, $p$, $d$, $f$, ... orbitals) and linear symmetry
($\sigma$, $\pi$, $\delta$, $\varphi$, ... orbitals), which are
likewise useful in relevant contexts.\citep{Lehtola2019_IJQC_25968,Lehtola2019_IJQC_25945,Lehtola2019_IJQC_25944,Lehtola2020_PRA_12516,Lehtola2023_JCTC_2502}
The implementation supports an arbitrary number of symmetry blocks,
which are defined by the calling program.

Each symmetry block for a given particle type typically has a different
size. Because all operations are independent for each type of particle
and symmetry block, the SCF problem is unambiguously defined in terms
of the individual symmetry blocks, each of which correspond to some
type of particle. The symmetry blocks thus form the elementary data
structure in the implementation.

Third, to account for various handlings of the spin, symmetry, and
differences in the ways occupations are handled for the various types
of particles (\emph{e.g.} in the NEO method), a maximal occupation
$f^{\max}$ must be defined separately for the orbitals in each block.
For instance, a spin-restricted calculation without symmetry ($C_{1}$
symmetry) fits 2 electrons per spatial orbital, while a spin-unrestricted
calculation fits 1 electron per spin-orbital. When occupations are
averaged,\citep{Lehtola2020_PRA_12516} an atomic orbital with angular
momentum $l$ can fit $2l+1$ electrons in a spin-unrestricted calculation,
and again double the amount in a spin-restricted calculation. The
abstraction of the maximum occupation allows updates to occupations
to be easily handled with a single routine. Defining the total number
of particles of each type, and the maximal orbital occupation in each
symmetry block, the Aufbau configuration can be formed by looping
over all particle types, and then distributing the specified number
of particles across the orbitals of all symmetry blocks for that type
of particle in increasing energy.

Fourth, although the mainstream of quantum chemistry is carried out
with real-valued orbitals, complex-valued orbitals are necessary in
some contexts. As already mentioned in the Introduction, orbitals
in general become complex in the presence of finite magnetic fields,\citep{Lehtola2020_MP_1597989,Aastroem2023_JPCA_10872}
and one of the main motivations in this work is to introduce reusable
software infrastructure that can also be used within this context.
Complex wave functions sometimes also arise in molecular problems,
as discussed by \citet{Small2015_JCP_24104}. 

Calculations with periodic boundary conditions are another stereotypical
example for the need of complex arithmetic, the problem here being
to solve Roothaan equations at every $\boldsymbol{k}$-point in the
first Brillouin zone\citep{Pisani1980_IJQC_501} 
\begin{equation}
\boldsymbol{F}^{\sigma}(\boldsymbol{k})\boldsymbol{C}^{\sigma}(\boldsymbol{k})=\boldsymbol{S}^{\sigma}(\boldsymbol{k})\boldsymbol{C}^{\sigma}(\boldsymbol{k})\boldsymbol{E}^{\sigma}(\boldsymbol{k}),\label{eq:roothaan-bz}
\end{equation}
where the Fock and overlap matrices are in general complex for $\boldsymbol{k}\neq\boldsymbol{0}$;
note here that $\boldsymbol{k}$ and $\sigma$ label symmetry blocks.
Since SCF acceleration in periodic calculations follows closely the
molecular approach,\citep{Maschio2018_TCA_60} and \ref{eq:roothaan-bz}
is still of the same form as \ref{eq:roothaan-nonorth}, the library
should also support periodic calculations without changes. If necessary,
the code is straightforward to amend to allow \textbf{k}-point weighting
in the convergence acceleration.\citep{Kresse1996_PRB_11169,Maschio2018_TCA_60}

Fully relativistic four-component calculations offer another example
of the need of complex-valued SCF, relativistic effects being famously
important for the chemistry of heavy elements.\citep{Pyykkoe2012_ARPC_45,Pyykkoe2012_CR_84}
The main complication in four-component calculations is the Dirac
sea of negative-energy solutions, which correspond to positronic solutions.\citep{Saue2011_C_94}
One should only populate the positive-energy orbitals that correspond
to electrons, and this is achieved by skipping over the negative-energy
solutions when determining the occupations of the four-component spinors.\citep{Saue2011_C_94}
The number of negative-energy solutions is given by the number of
(linearly independent) small-component basis functions in the calculation,
which is thus known beforehand. Although this feature does not exist
at present in OpenOrbitalOptimizer, the required change is thus trivial.

Note that relativistic effects are often described within approaches
that do not feature the Dirac sea, the negative energy states being
projected out by design. Popular approaches in present-day quantum
chemistry include the use of effective core potentials,\citep{Schwerdtfeger2011_C_3143}
and the use of the exact two-component (X2C) method.\citep{Heully1986_JPBAMP_2799,Kutzelnigg2005_JCP_241102,Liu2007_JCP_114107,Ilias2007_JCP_64102}
In contrast to the four-component calculations discussed in the previous
paragraph, these methods do not cause any extra complications, as
the SCF is carried out for these methods exactly in the same way as
in the non-relativistic case. 

Obviously, it is nowadays straightforward to develop generalized code
to handle both real and complex orbitals in a single implementation
without having to resort to any kind of code duplication by the use
of templated code. The implementations in OpenOrbitalOptimizer accept
a template data type for the orbitals as well as the orbital energies,
which means that the same code can be compiled in single or double
precision with real or complex orbitals, Fock, and density matrices.

\subsection{Convergence accelerators \label{subsec:Convergence-accelerators}}

The abstractions in \ref{sec:Theory} are a powerful tool for writing
reusable code. When any algorithm is implemented following these abstractions,
it should work regardless of the context it is eventually used in.
Next, we will briefly review the methods that represent the state
of the art in orbital optimization in the context of routine calculations.
As pointed out above, it suffices to discuss the algorithms only in
the context of a single symmetry block, regardless of the number of
types of particles involved in the calculation, and to accumulate
quantities over all symmetry blocks of all particle types. To make
the discussion clearer in the following, we therefore drop the particle
and symmetry block indices, and carry only the iteration index, as
all the operations now refer to individual symmetry blocks of an individual
type of particle.

\subsubsection{Pulay DIIS \label{subsec:Pulay-DIIS}}

We start with Pulay's DIIS method,\citep{Pulay1982_JCC_556} which
was introduced to speed up the convergence of the SCF equations and
is arguably still the most important technique used in electronic
structure calculations today.

To devise the metric for the DIIS error, let us examine the Fock matrix
in the orbital basis
\begin{equation}
\boldsymbol{F}^{\text{orb}}=\boldsymbol{C}^{\dagger}\boldsymbol{F}\boldsymbol{C}.\label{eq:fock-orb}
\end{equation}
If the orbitals are self-consistent, that is, \ref{eq:roothaan} applies,
the Fock matrix in the orbital basis given by \ref{eq:fock-orb} is
diagonal. In fact, it is easy to see that the off-diagonal elements
of the Fock matrix measure orbital gradients,\citep{Lehtola2020_M_1218}
and that the commutator of the Fock matrix $\boldsymbol{F}$ and the
density matrix $\boldsymbol{P}$ defined in \ref{eq:densmat}
\begin{equation}
\boldsymbol{e}=[\boldsymbol{F},\boldsymbol{P}]=\boldsymbol{F}\boldsymbol{P}-\boldsymbol{P}\boldsymbol{F}\label{eq:diis-commutator}
\end{equation}
measures the orbital error, vanishing when \ref{eq:roothaan} is satisfied.\citep{Pulay1982_JCC_556}
We note again here that \ref{eq:diis-commutator} must be computed
in the orthonormal basis,\citep{Pulay1982_JCC_556} which is the reason
why the choice to simplify the problem of \ref{eq:roothaan-nonorth}
to \ref{eq:std-eig-notilde} is not a limitation.

The DIIS error is given by the norm of the error vector
\begin{equation}
\epsilon^{\text{DIIS}}=\left\Vert \boldsymbol{e}\right\Vert .\label{eq:diis-error}
\end{equation}
Common choices for this norm in various programs are the root-mean-square
(rms) error
\begin{equation}
\epsilon^{\text{DIIS}}=\left\Vert \boldsymbol{e}\right\Vert _{\text{rms}}=\frac{1}{m}\sqrt{\sum_{i,j=1}^{m}|e_{ij}|^{2}}\label{eq:diis-error-rms}
\end{equation}
as well as the $L^{\infty}$ norm
\begin{equation}
\epsilon^{\text{DIIS}}=\left\Vert \boldsymbol{e}\right\Vert _{\infty}=\max_{ij}|e_{ij}|.\label{eq:diis-error-max}
\end{equation}
By default, we use the rms norm of \ref{eq:diis-error-rms}, which
appears to be the default in many programs; however, other choices
such as \ref{eq:diis-error-max} are also available in OpenOrbitalOptimizer.

The idea in DIIS is to determine an optimal mixing of the Fock matrices
$\boldsymbol{F}_{i}$ of previous iterations $i$
\begin{align}
\overline{\boldsymbol{F}} & =\sum_{i=1}^{m}c_{i}\boldsymbol{F}_{i},\label{eq:fock-mix}
\end{align}
in order to speed up the slow convergence of the iterative Roothaan
procedure. The parameter $m$ in \ref{eq:fock-mix} is the length
of the history. The length of the history is limited by default to
$m=10$ in OpenOrbitalOptimizer, following the default in Psi4.\citep{Smith2020_JCP_184108}
Note that there is a wide variation for the default value in various
programs, e.g. ORCA\citep{Neese2022_WIRCMS_1606} employs $m=5$,
Q-Chem\citep{Epifanovsky2021_JCP_84801} $m=15$, and Gaussian\citep{Frisch2009__a}
employs $m=20$ by default. 

The DIIS extrapolation weights are determined by minimizing the extrapolated
error
\begin{equation}
\overline{\boldsymbol{e}}=\sum_{i=1}^{m}c_{i}\boldsymbol{e}_{i}\label{eq:diis-extrapolated-error}
\end{equation}
with the condition that the weights sum up to one
\begin{equation}
\sum_{i=1}^{m}c_{i}=1.\label{eq:diis-sum}
\end{equation}
Minimization of $\left\Vert \overline{\boldsymbol{e}}\right\Vert ^{2}$
under the constraint of \ref{eq:diis-sum} yields the Lagrangian
\begin{equation}
L=\frac{1}{2}\boldsymbol{c}^{\text{T}}\boldsymbol{B}\boldsymbol{c}-\lambda\left(\sum_{i}c_{i}-1\right),\label{eq:DIIS-lagrangian}
\end{equation}
where $\lambda$ is a Lagrange multiplier, and 
\begin{equation}
B_{ij}=\langle\boldsymbol{e}_{i},\boldsymbol{e}_{j}\rangle\label{eq:diis-error-matrix}
\end{equation}
are elements of the DIIS error matrix. The Lagrangian is minimized
by weights that satisfy the linear system of equations\citep{Pulay1980_CPL_393}
\begin{equation}
\left(\begin{array}{cc}
\boldsymbol{B} & -\boldsymbol{1}\\
-\boldsymbol{1}^{\text{T}} & 0
\end{array}\right)\left(\begin{array}{c}
\boldsymbol{c}\\
\lambda
\end{array}\right)=\left(\begin{array}{c}
\boldsymbol{0}\\
-1
\end{array}\right).\label{eq:C1-DIIS}
\end{equation}

For large history sizes, the error vectors can become linearly dependent.
\citet{Hamilton1986_JCP_5728} suggested multiplying the diagonal
elements of \ref{eq:C1-DIIS} by a factor $1+d$ where $d=0.02$ is
a damping factor.

As the last row and column of \ref{eq:C1-DIIS} is always of the order
of unity, the numerical conditioning of the linear group of equations
can be improved by scaling the errors given by \ref{eq:diis-error-matrix}
such that the most recent diagonal element becomes unity.\citep{Eckert1997_JCC_1473}
In our implementation, the matrix is normalized with respect to the
smallest diagonal element $B_{ii}$, instead; this should have no
effect on the numerical behavior.

In addition, there is literature suggesting that limiting the matrices
included in DIIS extrapolation may help in some cases. We use the
simple criterion suggested by \citet{Chupin2021_EMMaNA_2785} for
restarting DIIS: only those vectors $i$ of the current history are
included in the extrapolation that satisfy
\begin{equation}
\delta^{\text{DIIS}}B_{ii}\leq\min_{j}B_{jj}\label{eq:diis-restart}
\end{equation}
with the default parameter value $\delta^{\text{DIIS}}=10^{-4}$.

\subsubsection{EDIIS and ADIIS}

DIIS can be viewed as a quasi-Newton method,\citep{Rohwedder2011_JMC_1889,Chupin2021_EMMaNA_2785}
and it generally leads to fast convergence when the orbitals are close
to convergence, \emph{i.e.}, when the orbital gradient is small. However,
this also means that DIIS can only be trusted to work when one is
close to a minimum; otherwise, the method can also easily converge
onto a saddle point. If the initial guess is poor, other methods must
therefore be used instead.

Energy DIIS\citep{Kudin2002_JCP_8255} (EDIIS) and augmented DIIS\citep{Hu2010_JCP_54109}
(ADIIS) are based on proxy energy functionals, which allow the methods
to work far from the optimum. The two methods are equivalent in the
case of Hartree--Fock theory,\citep{Garza2012_JCP_54110} but differ
in the case of density-functional theory\citep{Hohenberg1964_PR_864,Kohn1965_PR_1133}
(DFT) for which the energy is not a quadratic function of the density
matrix.

Both EDIIS and ADIIS are defined as interpolation methods, finding
an updated Fock matrix from \ref{eq:fock-mix} with the restriction
of \ref{eq:diis-sum} and that the weights be positive
\begin{equation}
c_{i}\geq0.\label{eq:aediis-pos}
\end{equation}
 EDIIS\citep{Kudin2002_JCP_8255} eponymously determines the weights
$c_{i}$ by minimizing the Hartree--Fock energy functional $E^{\text{HF}}$
for the linear combination of the density matrices
\begin{align}
E^{\text{EDIIS}}(\boldsymbol{c}) & =E^{\text{HF}}\left(\sum_{i=0}^{k}c_{i}\boldsymbol{P}_{i}\right)\label{eq:f-EDIIS-definition}\\
 & =\sum_{i=0}^{k}c_{i}E_{i}^{\text{SCF}}-\frac{1}{2}\sum_{i,j=0}^{k}\boldsymbol{c}\cdot\boldsymbol{X}\cdot\boldsymbol{c}\label{eq:E-EDIIS}
\end{align}
where $E^{\text{SCF}}$ is the total energy of calculation $i$, and
the matrix $\boldsymbol{X}$ is given by 
\begin{equation}
X_{ij}=\langle\boldsymbol{F}_{i}-\boldsymbol{F}_{j}|\boldsymbol{P}_{i}-\boldsymbol{P}_{j}\rangle.\label{eq:X-EDIIS}
\end{equation}
It is straightforward to show following the treatise of \citet{Kudin2002_JCP_8255}
that in the case of many kinds of particles, such as spin-up and spin-down
electrons in spin-polarized calculations, it suffices to accumulate
the $\boldsymbol{X}$ matrix over all types of particles (see proof
in Appendix).

ADIIS employs the quadratic energy function of \citet{Host2008_JCP_124106},
which yields the expression\citep{Hu2010_JCP_54109}
\begin{align}
E^{\text{ADIIS}}(\boldsymbol{c}) & =E(\boldsymbol{P}_{n})+2\sum_{i=1}^{n}c_{i}\langle\boldsymbol{P}_{i}-\boldsymbol{P}_{n}|\boldsymbol{F}_{n}\rangle\nonumber \\
 & +\sum_{i=1}^{n}\sum_{j=1}^{n}c_{i}c_{j}\langle\boldsymbol{P}_{i}-\boldsymbol{P}_{n}|\boldsymbol{F}_{j}-\boldsymbol{F}_{n}\rangle.\label{eq:E-ADIIS}
\end{align}

The coefficients $c_{i}$ are found by minimizing the corresponding
energy, \ref{eq:E-EDIIS} for EDIIS and \ref{eq:E-ADIIS} for ADIIS,
while imposing the constraints of \cref{eq:diis-sum,eq:aediis-pos}.
\citet{Hu2010_JCP_54109} proposed carrying out the minimization in
terms of unconstrained auxiliary variables $\boldsymbol{x}$ by implicit
satisfaction of both constraints
\begin{equation}
c_{i}=\frac{x_{i}^{2}}{\sum_{j}x_{j}^{2}},\label{eq:c-x}
\end{equation}
and such implementations appear to be widely used, e.g. in ERKALE,\citep{Lehtola2012_JCC_1572}
Psi4,\citep{Smith2020_JCP_184108} PySCF,\citep{Sun2020_JCP_24109}
and Q-Chem.\citep{Epifanovsky2021_JCP_84801} However, this is a bad
idea since the minimization problem is no longer convex in $\boldsymbol{x}$.
For example, we have found that this technique sometimes fails in
the line search. This issue is easily understood in the case where
the minimum features $c_{i}=0$ for some $c_{i}$: while the line
searches will try to make the ratios of the components of $\boldsymbol{x}$
larger, the resulting $\boldsymbol{c}$ parameters are invariant to
the scaling $\boldsymbol{x}\to\lambda\boldsymbol{x}$, which means
that the optimization becomes slowly convergent.

Since the problem is usually rather low-dimensional, being fully determined
by the length of the history, we will use a simple method inspired
by Powell's method,\citep{Powell1964_CJ_155} instead, which also
appears to have been used as a fallback by \citet{Kudin2002_JCP_8255}.

Both \cref{eq:E-EDIIS,eq:E-ADIIS} are quadratic forms of the type
\begin{equation}
f(\boldsymbol{x})=\frac{1}{2}\boldsymbol{x}^{\text{T}}\boldsymbol{A}\boldsymbol{x}+\boldsymbol{b}\cdot\boldsymbol{x}\label{eq:quadratic}
\end{equation}
with $\boldsymbol{A}=\boldsymbol{A}^{\text{T}}$. Because the models
of \ref{eq:quadratic} are small with a dimension defined by the DIIS
history length (values discussed after \ref{eq:C1-DIIS}), it does
not really matter if the algorithm used to minimize \ref{eq:quadratic}
is suboptimal, as long as it is robust, such that it always succeeds
in converging (close) to the minimum of \ref{eq:quadratic}.

Mathematically, the constraints of \cref{eq:diis-sum,eq:aediis-pos}
correspond to optimization on the unit simplex. The initial point
$\boldsymbol{x}_{0}$ is found by evaluating \ref{eq:quadratic} over
the edges $\boldsymbol{c}_{i}$, the center point $\boldsymbol{x}=(1/N,\dots,1/N)^{\text{T}}$,
as well as slightly offset points $\boldsymbol{x}=(3/(N+2),1/(N+2),\dots,1/(N+2))^{\text{T}}$.
The solution is then optimized starting from the point that gave the
lowest value of $f(\boldsymbol{x})$.

To ensure our solution is always in the allowed region, we update
$\boldsymbol{x}$ by pairwise searches
\begin{equation}
\boldsymbol{x}_{k+1}=(1-\lambda)\boldsymbol{x}_{k}+\lambda\boldsymbol{c}_{i}\label{eq:linesearch}
\end{equation}
where $(\boldsymbol{c}_{i})_{j}=\delta_{ij}$ corresponds to an individual
entry in the history. It is easy to show that the optimal value of
$\lambda$ for the line search of \ref{eq:linesearch} is 
\begin{equation}
\lambda=-\frac{\boldsymbol{c}_{i}^{\text{T}}\boldsymbol{A}\boldsymbol{x}+\boldsymbol{b}\cdot\boldsymbol{c}_{i}}{\boldsymbol{c}_{i}^{\text{T}}\boldsymbol{A}\boldsymbol{c}_{i}},\label{eq:optstep}
\end{equation}
where the allowed region is $0<\lambda\leq1$. Searches of the form
of \ref{eq:linesearch} are repeatedly looped over $i$, until the
value of $f(\boldsymbol{x})$ does not change appreciably any more.

\subsubsection{Hybrid scheme of EDIIS/ADIIS followed by DIIS}

Because EDIIS and ADIIS work well far from the optimum, while DIIS
works well close to it, we follow \citet{Garza2012_JCP_54110} and
interpolate between the two sets of weights
\begin{equation}
\boldsymbol{c}^{\text{hybrid}}=f^{\text{int}}(\epsilon^{\text{DIIS}})\boldsymbol{c}^{\text{DIIS}}+\left[1-f^{\text{int}}(\epsilon^{\text{DIIS}})\right]\boldsymbol{c}^{\text{EDIIS/ADIIS}}\label{eq:c-hybrid}
\end{equation}
such that 
\begin{itemize}
\item EDIIS/ADIIS is used exclusively when the DIIS error of \ref{eq:diis-error}
is large, $\epsilon^{\text{DIIS}}\ge\epsilon_{2}=10^{-1}$, 
\item DIIS is used exclusively when the DIIS error is small, $\epsilon^{\text{DIIS}}\le\epsilon_{1}=10^{-4}$,
and 
\item otherwise, a linear interpolating between the two is used.
\end{itemize}
Although the form \cref{eq:f-diis-interp} is not obvious in the equation
at the bottom of page 3 of \citeref{Garza2012_JCP_54110}, it appears
that \citet{Garza2012_JCP_54110} indeed used 
\begin{equation}
f^{\text{int}}(\epsilon^{\text{DIIS}})=\begin{cases}
0, & \epsilon\ge\epsilon_{2},\\
1-{\displaystyle \frac{\epsilon^{\text{DIIS}}-\epsilon_{2}}{\epsilon_{1}-\epsilon_{2}}}, & \epsilon_{1}<\epsilon<\epsilon_{2},\\
1, & \epsilon\le\epsilon_{1}.
\end{cases}\label{eq:f-diis-interp}
\end{equation}

We note that \citet{Garza2012_JCP_54110} suggested using pure EDIIS/ADIIS
if the error of the last cycle is 10\% greater than the minimum error.
We note that the error can also increase even when the energy is lowered,
and that we found this criterion to sometimes lead to convergence
to local minima. For this reason, we decided not to implement this
feature in OpenOrbitalOptimizer.

To choose between EDIIS and ADIIS in a given iteration of \ref{eq:c-hybrid},
we employ the minimal error sampling algorithm (MESA) of \citet{GarciaChavez2018__}:
we compute both the EDIIS and ADIIS extrapolations, and pick the one
that results in the least change to the density matrix as measured
by the Frobenius norm $\left\Vert \boldsymbol{P}_{\text{current}}-\boldsymbol{P}_{\text{new}}\right\Vert _{F}$.

\subsubsection{History cleanup}

Even though ADIIS and EDIIS are interpolation methods that are based
on solid mathematical models, using faraway points can still lead
to degraded performance in DFT calculations: our exploratory calculations
showed that when the orbitals undergo large changes in occupations,
it is beneficial to clear the history. In our implementation, following
the idea of \citet{Chupin2021_EMMaNA_2785} discussed above in \ref{eq:diis-restart},
we keep only those density matrices in the history that satisfy 
\begin{equation}
\delta^{\text{history}}\left\Vert \boldsymbol{P}_{i}-\boldsymbol{P}_{0}\right\Vert _{\text{F}}\leq\min_{j\in[1,\dots,N-1]}\left\Vert \boldsymbol{P}_{j}-\boldsymbol{P}_{0}\right\Vert _{\text{F}}\label{eq:history-cleanup}
\end{equation}
where $\boldsymbol{P}_{0}$ is the density matrix that led to the
lowest energy, the parameter $\delta$ again has the default value
$\delta^{\text{history}}=10^{-4}$, and $\left\Vert \right\Vert _{F}$
denotes the Frobenius norm.

\subsubsection{Optimal damping}

To continue the above discussion, when a poor initial guess is used
(\emph{e.g.}, the core Hamiltonian for a heavy atom), the orbital
gradient can be huge, and the changes in the density will be large.
In this case, even the ADIIS/EDIIS procedure outlined above can fail.
The problem here is that some sort of restart is necessary to make
ADIIS/EDIIS work with DFT methods, but when large changes in the density
matrix occur, the restart may be too aggressive, cleaning up all of
the preceding history.

We have solved this issue by using the optimal damping algorithm (ODA)
of \citet{Cances2000_IJQC_82} in the case of large maximum orbital
gradients measured by \ref{eq:diis-error-max}, which is an intensive
criterion, independent of system size.

In the case of a large maximum orbital gradient, the default threshold
for which we set as $\left\Vert \epsilon\right\Vert _{\infty}\ge1$,
our procedure to relax the density matrix is as follows. In the case
of a single type of particle (as in spin-restricted Hartree--Fock
or DFT, for example), the relaxation problem is one-dimensional: we
carry out a line search of the energy in terms of a mixture of two
density matrices
\begin{equation}
E(\lambda)=E[(1-\lambda)\boldsymbol{P}_{0}+\lambda\boldsymbol{P}_{1}],\label{eq:oda}
\end{equation}
where $\boldsymbol{P}_{0}$ is again the density matrix that led to
the lowest energy, and $\boldsymbol{P}_{1}$ is the density matrix
built from the orbitals that diagonalize $\boldsymbol{F}_{0}=\boldsymbol{F}(\boldsymbol{P}_{0})$.
We perform a cubic interpolation to find $\lambda$ in \ref{eq:oda}.\citep{Cances2001_JCP_10616}

If $E(\boldsymbol{P}_{1})<E(\boldsymbol{P}_{0})$, we accept the step.
Otherwise, we already know $E(0)=E(\boldsymbol{P}_{0})=E_{0}$ and
$\boldsymbol{F}_{0}=\boldsymbol{F}(\boldsymbol{P}_{0})$, as well
as the values at $\lambda=1$, $E(1)=E(\boldsymbol{P}_{1})=E_{1}$
and $\boldsymbol{F}_{1}=\boldsymbol{F}(\boldsymbol{P}_{1})$. The
derivatives with respect to $\lambda$ are given by these data as
$E'(0)=\text{Tr }\boldsymbol{F}_{0}(\boldsymbol{P}_{1}-\boldsymbol{P}_{0})$
and $E'(1)=\text{Tr }\boldsymbol{F}_{1}(\boldsymbol{P}_{1}-\boldsymbol{P}_{0})$.
The four data points $E(0)$, $E'(0)$, $E(1)$ and $E(1)$ afford
a cubic fit.

If the cubic fit predicts a minimum at $0<\lambda^{*}<1$ (if there
are two minima in the allowed region we pick the lower one predicted
by the fit), we compute the corresponding natural orbitals and orbital
occupations
\begin{equation}
(1-\lambda^{*})\boldsymbol{P}_{0}+\lambda^{*}\boldsymbol{P}_{1}=\boldsymbol{C}\boldsymbol{f}\boldsymbol{C}^{\text{T}}\label{eq:oda-dens}
\end{equation}
and evaluate the corresponding total energy and Fock matrix. 

If there are several types of particles, the optimization problem
is many-dimensional. For example, for the case of spin-unrestricted
Hartree--Fock or density-functional theory, we have to optimize
\begin{equation}
E(\lambda_{\alpha,}\lambda_{\beta})=E[\boldsymbol{P}^{\alpha}(\lambda_{\alpha}),\boldsymbol{P}^{\beta}(\lambda_{\beta})]\label{eq:oda-2d}
\end{equation}
in the unit square $\boldsymbol{\lambda}=(\lambda_{\alpha},\lambda_{\beta})\in[0,1]^{2}\backslash(0,0)$,
with separate interpolations of the form \ref{eq:oda-dens} for either
spin. Knowing the current Fock matrices $\boldsymbol{F}_{0}^{\sigma}$
for both spins, we can build the new density matrices $\boldsymbol{P}_{1}^{\sigma}$,
and compute the energy gradients with respect to the mixing factors
\begin{equation}
\partial E/\partial\lambda_{\sigma}=\text{Tr }\boldsymbol{F}_{0}^{\sigma}(\boldsymbol{P}_{1}^{\sigma}-\boldsymbol{P}_{0}^{\sigma}).\label{eq:oda-gradient}
\end{equation}
The negative gradient $-\nabla_{\boldsymbol{\lambda}}E$ then determines
the search direction. Since the search always starts from the current
solution, which corresponds to $\boldsymbol{\lambda}=\boldsymbol{0}$,
the components of the search direction $\boldsymbol{s}$ are obtained
as
\begin{equation}
s_{i}=\max(0,-\partial E/\partial\lambda_{i})\label{eq:oda-direction}
\end{equation}
to ensure we are in the allowed region. The trial $\boldsymbol{\lambda}^{\text{trial}}$
is then obtained by stepping to the boundary,
\begin{equation}
\boldsymbol{\lambda}^{\text{trial}}=\frac{\boldsymbol{s}}{\max_{i}s_{i}}.\label{eq:lambda-trial}
\end{equation}
The energy $E(\boldsymbol{\lambda}^{\text{tr}})$ is then evaluated,
and again if $E(\boldsymbol{\lambda}^{\text{tr}})<E(\boldsymbol{0})$
the step is accepted. Otherwise, we perform a cubic interpolation
along $\boldsymbol{\lambda}^{\text{tr}}$ exactly as in the case of
one dimension. Note that even though the example discussed herein
only had two types of particles, this procedure is valid for an arbitrary
number of types of particles. 

ODA steps are repeated until the maximum orbital gradient becomes
acceptable. The history generated by the ODA algorithm may then be
used by the ADIIS/EDIIS interpolation and DIIS extrapolation methods.

\section{Design Choices \label{sec:Design-Choices}}

Arguably the most important design choice in any project is the license.
In order to promote the reusability of the library and avoid the need
for duplicate solutions,\citep{Lehtola2023_JCP_180901} we chose to
license OpenOrbitalOptimizer under the Mozilla Public License that
is also used in Libxc,\citep{Lehtola2018_S_1} which is nowadays used
by over 40 programs including also several commercial programs.

We chose to make OpenOrbitalOptimizer a header-only library in order
to facilitate its inclusion in various programs. The language was
chosen as C++ as the standard in scientific computing; there are many
recent quantum chemistry codes in C++, and the C++ language makes
templating easy.

Moreover, as the main tasks of the library have to do with linear
algebra, the library is developed on top of the Armadillo library,\citep{Sanderson2016_JOSS_26}
which offers a convenient syntax for matrix and vector operations,
as well as the required support for templated datatypes that are used
throughout OpenOrbitalOptimizer. The use of Armadillo within the library
does not restrict the use of the library in other codes, since the
matrix data is easy to convert to other formats in the library interface.

The main design decision in OpenOrbitalOptimizer is the limitation
to problem sizes that are most common in quantum chemistry, that is,
matrix sizes that can be stored in core memory, and diagonalized in
full. This is a conscious design decision, motivated by the want to
have a lean library that is easy to work with, and easy to include
in other codes. The library can still be potentially useful in other
contexts, such as calculations in plane-wave basis sets, as one can
always find a smaller orbital basis to use OpenOrbitalOptimizer in,
for example by computing the $N$ lowest crystalline or molecular
orbitals for a guess Hamiltonian by iterative diagonalization.

Since evaluation of the Fock matrix can be more efficient in terms
of orbital coefficients and occupation numbers rather than a density
matrix (this is the case with resolution-of-the-identity techniques
for exact exchange, as well as DFT in large orbital basis sets, for
example), the basic data that OpenOrbitalOptimizer works with are
the molecular orbital coefficients $\boldsymbol{C}$ and occupation
numbers $\boldsymbol{f}$ for all symmetry blocks.

The library is used by defining the numbers of blocks per particle
type (which also implicitly defines the number of particle types),
the maximum occupation in each block, the number of particles of each
type, a function to evaluate the Fock matrix and total energy from
the given $\boldsymbol{C}$ and $\boldsymbol{f}$, and the descriptions
of each of the blocks. Calculations can be initialized either by supplying
initial guesses for $\boldsymbol{C}$ and $\boldsymbol{f}$, or by
supplying a guess for the Fock matrix $\boldsymbol{F}$.

In our experience, the SCF implementations in many programs can converge
onto a higher energy than what was encountered during the SCF iterations.
A key goal in our implementation is to avoid this by never erasing
the lowest-energy estimate for the solution. We always keep the lowest-energy
solution first in our stack, after which the iterates are sorted in
decreasing iteration number. When the stack is full, we erase the
oldest non-minimal solution, which is kept at the end of the stack.

This procedure has the potential pitfall in that it can get stuck.
We decided to resolve this issue by triggering an alternative strategy
in the case $N/2$ consecutive (A/E)DIIS iterations have failed to
result in a decrease of the total energy, $N$ being the maximum allowable
history length: our tit-for-tat strategy is to carry out the next
$N/2$ iterations with ODA in the hopes to make sure that the solution
gets sufficiently far away from the stagnant region for (A/E)DIIS
to recover its usefulness.

\section{Computational Details\label{sec:Computational-Details}}

Although OpenOrbitalOptimizer has already been interfaced with several
programs, such as HelFEM\citep{Lehtola2019_IJQC_25944,Lehtola2019_IJQC_25945,HelFEM}
and Psi4,\citep{Smith2020_JCP_184108} we exemplify the library with
nuclear-electronic orbital\citep{Webb2002_JCP_4106,Pavosevic2020_CR_4222,HammesSchiffer2021_JCP_30901}
(NEO) Hartree--Fock (HF) calculations that we implemented in a plugin
to the ERKALE program.\citep{Lehtola2012_JCC_1572,Lehtola2018__a}

NEO-HF consists of solving a coupled set of equations for the protons
and electrons, which read in the atomic-orbital basis as\citep{Webb2002_JCP_4106}
\begin{equation}
\begin{cases}
\boldsymbol{F}^{\textrm{e}}\boldsymbol{C}^{\textrm{e}}=\boldsymbol{S}^{\textrm{e}}\boldsymbol{C}^{\textrm{e}}\boldsymbol{P}^{\textrm{e}},\\
\boldsymbol{F}^{\textrm{p}}\boldsymbol{C}^{\textrm{p}}=\boldsymbol{S}^{\textrm{p}}\boldsymbol{C}^{\textrm{p}}\boldsymbol{P}^{\textrm{p}}.
\end{cases}\label{eq:neo-roothaan}
\end{equation}
The electronic and protonic Fock matrices are given by
\begin{align}
\boldsymbol{F}^{\text{e}} & =\boldsymbol{T}^{\text{e}}+\boldsymbol{V}^{\text{e}}+\boldsymbol{J}^{\text{ee}}+\boldsymbol{K}^{\text{ee}}+\boldsymbol{J}^{\text{pe}}\label{eq:el-fock}\\
\boldsymbol{F}^{\text{p}} & =\boldsymbol{T}^{\text{p}}+\boldsymbol{V}^{\text{p}}+\boldsymbol{J}^{\text{pp}}+\boldsymbol{K}^{\text{pp}}+\boldsymbol{J}^{\text{ep}}\label{eq:p-fock}
\end{align}
where $\boldsymbol{T}^{\text{e}}$ is the electronic kinetic energy
matrix, $\boldsymbol{V}^{\text{e}}$ are the nuclear attraction integrals
for the Born--Oppenheimer nuclei, $\boldsymbol{J}^{\textrm{ee}}$
is the electron-electron Coulomb matrix, $\boldsymbol{K}^{\text{ee}}$
is the electron-electron exchange matrix, and $\boldsymbol{J}^{\text{pe}}$
is the proton-electron Coulomb matrix, \emph{i.e.}, the Coulomb potential
that the protons generate for the electrons. As usual in NEO, the
protons are assumed to be in a high-spin state in all calculations. 

As electronic basis sets, we consider the polarization consistent
pc-n and aug-pc-n basis sets of Jensen.\citep{Jensen2001_JCP_9113,Jensen2002_JCP_9234}
Following our recent work in \citeref{Nikkanen2025__}, we uncontract
the electronic basis set on the quantum protons, as this leads to
significant improvements in the precision of the calculation; this
procedure is denoted as uncHq-. The protonic basis set is PB4-F1\citep{Yu2020_JCP_244123}
in all calculations. All basis sets are taken from the Basis Set Exchange.\citep{Pritchard2019_JCIM_4814}

We use density fitting in all calculations with autogenerated auxiliary
basis sets. As recently reviewed by \citet{Pedersen2023_WIRCMS_1692},
there is a close connection between density fitting and Cholesky decompositions.
\citet{Liu2023_JCTC_6255} showed that the Cholesky decomposition
yields good results in NEO calculations when the decomposition is
carried out for the union of the protonic and electronic basis functions;
we thus generated the auxiliary basis set from pivoted Cholesky decompositions
for the union of the protonic and electronic basis functions following
our recent work.\citep{Lehtola2021_JCTC_6886,Lehtola2023_JCTC_6242}
We then used the same auxiliary basis set for both the electrons and
the protons.

We assess the accuracy of our density-fitted NEO-HF implementation
against calculations performed with exact integrals in Q-Chem.\citep{Epifanovsky2021_JCP_84801}
The two-electron integral threshold was tightened to $\epsilon=10^{-12}$
in Q-Chem. The basis set linear dependence thresholds were set to
$10^{-5}$ in both codes. The Coulomb overlap matrix in the density
fitting approach was inverted using the pivoted Cholesky technique
of \citeref{Lehtola2019_JCP_241102}, using a $10^{-8}$ threshold
to choose the pivot functions, and a $10^{-7}$ linear dependence
threshold for the canonical orthonormalization step.

Solving \ref{eq:neo-roothaan} requires simultaneous initial guesses
for both the electrons and the protons. A reasonable guess for the
electrons is obtained from standard atomic guesses, as proton delocalization
effects are likely localized in the electronic structure; we used
a superposition of atomic potentials\citep{Lehtola2019_JCTC_1593,Lehtola2020_JCP_144105}
for the present calculations. However, we found that guessing the
protonic Fock matrix based on the electronic solution leads to poor
results.

Instead, we found that a minimal 1s protonic basis guess with exponent
$16\sqrt{2}\approx22.627$ adopted from the 8s8p8d even-tempered basis
of \citet{Yang2017_JCP_114113} leads to qualitatively correct electronic
and protonic orbitals, as calculations in the larger basis sets with
this read-in guess converge to the DIIS region in just a few iterations.
Performing this initial calculation has practically no overhead, since
there are no protonic degrees of freedom to optimize in the minimal
basis.

We note that \citet{Liu2022_JPCA_7033} described using a guess where
the proton occupies the tightest protonic basis function for each
quantum nucleus---which we assume to refer to an s function. Our
guess is thus somewhat similar to that of \citeauthor{Liu2022_JPCA_7033},
but our guess is independent of the protonic basis used in the actual
calculation. Note that the tightest basis function in the protonic
8s8p8d basis of \citet{Yang2017_JCP_114113} has an exponent of 32.0,
while the tightest s function in PB4-F1\citep{Yu2020_JCP_244123}
has an exponent 28.950.

Highlighting the robustness of the reusable SCF solver in OpenOrbitalOptimizer,
NEO-HF was implemented in both the simultaneous and stepwise fashions.
In the stepwise approach, each macroiteration consists of first solving
the electronic problem at fixed proton-electron Coulomb potential,
then updating the electron-proton Coulomb potential, and then solving
the protonic problem in this fixed external Coulomb potential. The
macroiteration terminates by the update of the proton-electron Coulomb
potential. The energy is evaluated in full at every iteration by adding
in the quantum mechanical energy for the fixed particles (protons
for electrons, and electrons for protons).

Due to the differences in Fock matrix constructions, the simultaneous
and stepwise approaches can have very different costs per iteration,
as illustrated by \ref{tab:Terms-to-calculate}. Even though the simultaneous
optimization has been found to converge faster,\citep{Liu2022_JPCA_7033}
the electronic basis set is much larger than the protonic one in typical
calculations; this is especially the case when one is interested in
going towards the complete basis set limit, see \citet{Khan2025_JCC_70082}
and our recent work in \citeref{Nikkanen2025__}. Even though the
protonic problem tends to be more difficult to converge, requiring
many more iterations than the electronic one, the protonic iterations
are usually extremely cheap in comparison to the electronic ones.

We found the speed of convergence to be greatly influenced by the
DIIS history length in exploratory calculations. A history length
of 20 matrices were used in all the calculations reported herein;
going higher did not appear to lead to improved convergence.

\begin{table*}
\begin{centering}
\begin{tabular}{c|ll|l}
 & \multicolumn{2}{l|}{stepwise} & \tabularnewline
matrices & electronic & protonic & simultaneous\tabularnewline
\hline 
\hline 
$\boldsymbol{J}^{\textrm{ee}}$, $\boldsymbol{K}^{\textrm{ee}}$ & yes & no & yes\tabularnewline
$\boldsymbol{J}^{\textrm{pp}}$, $\boldsymbol{K}^{\textrm{pp}}$ & no & yes & yes\tabularnewline
$\boldsymbol{J}^{\textrm{ep}}$ & after convergence & no & yes\tabularnewline
$\boldsymbol{J}^{\textrm{pe}}$ & no & after convergence & yes\tabularnewline
\end{tabular}
\par\end{centering}
\caption{Terms to calculate in every iteration in NEO-HF. Each step of the
electronic (protonic) problem only requires computing the corresponding
exchange and Coulomb matrices, while the electron-proton (proton-electron)
Coulomb term is only evaluated at convergence. \label{tab:Terms-to-calculate}}

\end{table*}

\section{Results \label{sec:Results}}

We consider NEO-HF calculations on the protonated water tetramers
\ce{H3O+(H2O)3} with the four geometries used by \citet{Liu2022_JPCA_7033},
which were originally determined by \citet{Heindel2018_JCTC_4553}.
The geometries correspond to ring, eigen, ciszundel, and transzundel
structures. As all nine protons were modeled quantum mechanically,
these calculations are a good check for the robustness of the convergence
accelerators. All calculations were converged to a root-mean-square
(rms) orbital gradient, \emph{i.e.}, DIIS error of $10^{-7}$.

Calculations were carried out in both the simultaneous and stepwise
procedures. The stepwise procedure was carried out until the energy
changed less than $10^{-7}$ $\textrm{E}_{\textrm{h}}$ in a macroiteration,
at which point the calculations switched to the simultaneous algorithm;
in contrast, calculations using the simultaneous approach employ that
algorithm from the start.

Calculations were furthermore performed with auxiliary basis sets
of three sizes: the ``small'', ``large'', and ``verylarge''
auxiliary basis sets generated with the procedure of \citeref{Lehtola2023_JCTC_6242}.
As previous experience shows that at least a triple-$\zeta$ orbital
basis is required for generating reliable auxiliary basis sets,\citep{Lehtola2021_JCTC_6886}
data presented for the double-$\zeta$ pc-1 and aug-pc-1 orbital basis
sets employ the auxiliary basis set of the corresponding triple-$\zeta$
orbital basis set. 

\begin{table*}
\begin{centering}
\begin{tabular}{cc|c|rrrr}
Basis & Geometry & $N_{\text{sim}}$ & $N_{\textrm{elec}}$ & $N_{\textrm{prot}}$ & $N_{\text{ep/pe}}$ & $\Delta E$\tabularnewline
\hline 
\hline 
uncHq-pc-\textbf{1} & ring & 124 & 93 & 298 & 32 & -4.6e-08\tabularnewline
uncHq-pc-\textbf{1} & eigen & 123 & 91 & 276 & 35 & -7.2e-08\tabularnewline
uncHq-pc-\textbf{1} & ciszundel & 127 & 91 & 281 & 34 & -7.7e-08\tabularnewline
uncHq-pc-\textbf{1} & transzundel & 126 & 92 & 285 & 35 & -8.6e-08\tabularnewline
uncHq-\textbf{aug}-pc-\textbf{1} & ring & 108 & 69 & 229 & 21 & -2.8e-08\tabularnewline
uncHq-\textbf{aug}-pc-\textbf{1} & eigen & 100 & 60 & 210 & 12 & -1.0e-08\tabularnewline
uncHq-\textbf{aug}-pc-\textbf{1} & ciszundel & 99 & 65 & 205 & 20 & -3.4e-08\tabularnewline
uncHq-\textbf{aug}-pc-\textbf{1} & transzundel & 99 & 65 & 206 & 20 & -3.4e-08\tabularnewline
uncHq-pc-\textbf{2} & ring & 134 & 105 & 342 & 44 & -1.0e-07\tabularnewline
uncHq-pc-\textbf{2} & eigen & 132 & 105 & 329 & 47 & -1.3e-07\tabularnewline
uncHq-pc-\textbf{2} & ciszundel & 141 & 96 & 339 & 35 & -1.1e-07\tabularnewline
uncHq-pc-\textbf{2} & transzundel & 132 & 104 & 346 & 43 & -9.7e-08\tabularnewline
uncHq-\textbf{aug}-pc-\textbf{2} & ring & 113 & 69 & 292 & 17 & -7.3e-08\tabularnewline
uncHq-\textbf{aug}-pc-\textbf{2} & eigen & 111 & 68 & 290 & 16 & -6.3e-08\tabularnewline
uncHq-\textbf{aug}-pc-\textbf{2} & ciszundel & 114 & 69 & 299 & 17 & -5.5e-08\tabularnewline
uncHq-\textbf{aug}-pc-\textbf{2} & transzundel & 112 & 70 & 294 & 17 & -7.3e-08\tabularnewline
uncHq-pc-\textbf{3} & ring & 424 & 211 & 946 & 54 & -3.0e-07\tabularnewline
uncHq-pc-\textbf{3} & eigen & 532 & 183 & 836 & 48 & 2.0e-07\tabularnewline
uncHq-pc-\textbf{3} & ciszundel & 476 & 211 & 966 & 55 & -1.0e-06\tabularnewline
uncHq-pc-\textbf{3} & transzundel & 361 & 188 & 831 & 48 & -5.9e-08\tabularnewline
uncHq-\textbf{aug}-pc-\textbf{3} & ring & 379 & 201 & 942 & 54 & 2.7e-07\tabularnewline
uncHq-\textbf{aug}-pc-\textbf{3} & eigen & 444 & 173 & 837 & 48 & 1.8e-06\tabularnewline
uncHq-\textbf{aug}-pc-\textbf{3} & ciszundel & 510 & 195 & 950 & 54 & -6.7e-08\tabularnewline
uncHq-\textbf{aug}-pc-\textbf{3} & transzundel & 282 & 179 & 842 & 49 & 1.3e-06\tabularnewline
\end{tabular}
\par\end{centering}
\caption{Number of Fock matrix builds needed to converge the NEO-HF calculations
with the \textquotedblleft large\textquotedblright{} auxiliary basis
for various geometries of the protonated water tetramer: $N_{\textrm{sim}}$
is the number of Fock builds in the simultaneous electronic-protonic
SCF, while $N_{\text{elec}}$ and $N_{\text{prot}}$ are the total
numbers of electronic and protonic Fock builds in the stepwise SCF
procedure (\emph{c.f.} \ref{tab:Terms-to-calculate}). $N_{\textrm{ep/pe}}$
shows the number of times the electron-proton and proton-electron
coupling terms were calculated, while $\Delta E=E(\text{simultaneous})-E(\text{separate})$
shows the difference in the final total energies. \label{tab:Number-of-Fock}}
\end{table*}

The number of Fock matrix builds required in the various calculations
is shown in \ref{tab:Number-of-Fock}. As the table demonstrates,
the total number of Fock matrix builds is indeed smaller in the simultaneous
mode, as \citet{Liu2022_JPCA_7033} found. However, the stepwise solution
requires roughly a third fewer electronic Fock matrix builds than
the simultaneous solution; the electron-proton and proton-electron
Coulomb matrices also need to be built a considerably smaller number
of times. Even though the stepwise solution requires 2--3 times more
protonic Fock matrices, the protonic basis is typically a small fraction
of the size of the electronic basis, and these builds are extremely
cheap.

Furthermore, we find that the simultaneous solution slows down considerably
when going to larger electronic basis sets, and that the contrast
to the stepwise algorithm becomes even larger.

We assess the precision of the calculations with the three sizes of
auxiliary basis sets in \ref{tab:Precision-of-the}. In the larger
orbital basis sets, the fit errors in the total NEO-HF energies are
just a few $\upmu\textrm{E}_{\textrm{h}}$, causing no concern about
the accuracy of the calculations. The fitting errors in the total
energy are slightly larger in the calculations with the smallest orbital
basis sets, but even the error of $9\times10^{-5}\textrm{E}_{\textrm{h}}\approx0.06$
kcal/mol which is negligible. Note also that the error is very systematic
across the geometries, and will therefore exhibit systematic error
cancellation in computing isomer or reaction energies, for instance.

Clear improvement in precision can be observed for the smaller orbital
basis sets when using larger and larger auxiliary basis sets, but
this pattern is not so evident in the largest calculations, which
we interpret as an issue with finite numerical precision, as the large
auxiliary basis sets start to exhibit linearly dependencies.

\begin{table*}
\centering{}%
\begin{tabular}{ll|rrrr}
Basis & Model & eigen & transzundel & ciszundel & ring\tabularnewline
\hline 
\hline 
uncHq-pc-\textbf{1}$^{*}$ & Q-Chem & -304.0829921 & -304.0763508 & -304.0760041 & -304.0739284\tabularnewline
uncHq-pc-\textbf{1}$^{*}$ & TW, ``small'' & 0.0000922 & 0.0000922 & 0.0000922 & 0.0000939\tabularnewline
uncHq-pc-\textbf{1}$^{*}$ & TW, ``large'' & 0.0000561 & 0.0000560 & 0.0000560 & 0.0000574\tabularnewline
uncHq-pc-\textbf{1}$^{*}$ & TW, ``verylarge'' & 0.0000207 & 0.0000219 & 0.0000221 & 0.0000239\tabularnewline
uncHq-pc-\textbf{2} & Q-Chem & -304.2805872 & -304.2746806 & -304.2743526 & -304.2705210\tabularnewline
uncHq-pc-\textbf{2} & TW, ``small'' & 0.0000128 & 0.0000129 & 0.0000128 & 0.0000152\tabularnewline
uncHq-pc-\textbf{2} & TW, ``large'' & 0.0000010 & 0.0000010 & 0.0000010 & 0.0000020\tabularnewline
uncHq-pc-\textbf{2} & TW, ``verylarge'' & -0.0000030 & -0.0000016 & -0.0000015 & -0.0000006\tabularnewline
uncHq-pc-\textbf{3} & Q-Chem & -304.3062376 & -304.3003964 & -304.3001021 & -304.2963777\tabularnewline
uncHq-pc-\textbf{3} & TW, ``small'' & 0.0000039 & 0.0000031 & 0.0000027 & 0.0000033\tabularnewline
uncHq-pc-\textbf{3} & TW, ``large'' & 0.0000007 & 0.0000004 & -0.0000001 & 0.0000007\tabularnewline
uncHq-pc-\textbf{3} & TW, ``verylarge'' & -0.0000002 & -0.0000010 & -0.0000008 & -0.0000003\tabularnewline
uncHq-\textbf{aug}-pc-\textbf{1}$^{*}$ & Q-Chem & -304.0967476 & -304.0902633 & -304.0904477 & -304.0896750\tabularnewline
uncHq-\textbf{aug}-pc-\textbf{1}$^{*}$ & TW, ``small'' & 0.0000628 & 0.0000629 & 0.0000628 & 0.0000628\tabularnewline
uncHq-\textbf{aug}-pc-\textbf{1}$^{*}$ & TW, ``large'' & 0.0000232 & 0.0000225 & 0.0000229 & 0.0000220\tabularnewline
uncHq-\textbf{aug}-pc-\textbf{1}$^{*}$ & TW, ``verylarge'' & 0.0000223 & 0.0000222 & 0.0000220 & 0.0000204\tabularnewline
uncHq-\textbf{aug}-pc-\textbf{2} & Q-Chem & -304.2820253 & -304.2761474 & -304.2758457 & -304.2721887\tabularnewline
uncHq-\textbf{aug}-pc-\textbf{2} & TW, ``small'' & 0.0000046 & 0.0000047 & 0.0000048 & 0.0000051\tabularnewline
uncHq-\textbf{aug}-pc-\textbf{2} & TW, ``large'' & -0.0000016 & -0.0000027 & -0.0000027 & -0.0000027\tabularnewline
uncHq-\textbf{aug}-pc-\textbf{2} & TW, ``verylarge'' & -0.0000021 & -0.0000023 & -0.0000025 & -0.0000039\tabularnewline
uncHq-\textbf{aug}-pc-\textbf{3} & Q-Chem & -304.3064263 & -304.3005800 & -304.3002842 & -304.2965622\tabularnewline
uncHq-\textbf{aug}-pc-\textbf{3} & TW, ``small'' & 0.0000016 & 0.0000008 & 0.0000003 & 0.0000004\tabularnewline
uncHq-\textbf{aug}-pc-\textbf{3} & TW, ``large'' & -0.0000004 & 0.0000003 & 0.0000003 & 0.0000007\tabularnewline
uncHq-\textbf{aug}-pc-\textbf{3} & TW, ``verylarge'' & -0.0000022 & -0.0000009 & -0.0000034 & -0.0000014\tabularnewline
\end{tabular}\caption{Precision of the RI-NEO-HF calculations of this work (TW), compared
to analogous calculations carried out with traditional two-electron,
two-proton, electron-proton, and proton-electron integrals with Q-Chem.
The data are shown as deviations from the Q-Chem total energy. All
values are in $\mathrm{E}_{\mathrm{h}}$. $^{*}$Calculation employed
auxiliary basis set generated for the corresponding triple-$\zeta$
basis: uncHq-pc-2 or uncHq-aug-pc-2. \label{tab:Precision-of-the}}
\end{table*}

\section{Summary and Conclusions \label{sec:Summary-and-Conclusions}}

We have presented OpenOrbitalOptimizer,\citep{Lehtola2025__} a reusable
open-source header-only templated C++ library for orbital optimization
and self-consistent field (SCF) calculations. OpenOrbitalOptimizer
aims to either fully eliminate existing disparate SCF implementations
in various programs, or alternatively to supplement their functionalities
by offering access to consistent implementations of various SCF methodologies.
Attesting to the power of modern programming paradigms, including
extensive comments, the present implementation is only around 2400
lines of C++ code.

OpenOrbitalOptimizer offers a unified interface for solving coupled
SCF equations of an arbitrary number of types of particles, with an
arbitrary number of symmetries. OpenOrbitalOptimizer has already been
interfaced with HelFEM,\citep{HelFEM,Lehtola2019_IJQC_25944,Lehtola2019_IJQC_25945,Lehtola2020_MP_1597989,Lehtola2020_PRA_12516,Lehtola2023_JCTC_2502}
ERKALE,\citep{Lehtola2012_JCC_1572,Lehtola2018__a} and Psi4,\citep{Smith2020_JCP_184108}
and we hope to see many more interfaces with other programs. While
none of the currently interfaced codes perform fully relativistic
or solid-state calculations, the necessary extensions discussed in
the Introduction should be straightforward to make in OpenOrbitalOptimizer.

OpenOrbitalOptimizer is openly available free of charge under the
Mozilla Public License, with the main project hosted at \url{https://github.com/susilehtola/OpenOrbitalOptimizer}.
It runs on Linux, MacOS, and Windows.

The development of OpenOrbitalOptimizer was greatly motivated by the
needs to support nuclear-electronic orbital (NEO) calculations,\citep{Webb2002_JCP_4106,Pavosevic2020_CR_4222,HammesSchiffer2021_JCP_30901}
which feature also quantum protons in addition to electrons. We exemplified
OpenOrbitalOptimizer with NEO Hartree--Fock calculations on protonated
water clusters, and described a simple but powerful initial guess
for the protons. Calculations were carried out both with simultaneous
electron-proton SCF, as well as the stepwise SCF procedure consisting
of alternating SCF procedures for the electrons and protons, respectively.
We found the stepwise SCF procedure to exhibit stabler convergence.

In future work, we aim to extend the functionality of OpenOrbitalOptimizer
for calculations on systems with degeneracies at the Fermi level,
in which case the Aufbau theorem does not hold and the occupations
of the degenerate orbitals must instead be explicitly optimized. Although
our main interest in this problem arises from atomic calculations,\citep{Slater1969_PR_672,Lehtola2019_JCTC_1593,Lehtola2020_PRA_12516,Lehtola2023_JCTC_2502,Aastroem2025_JPCA_2791}
degenerate occupations can sometimes be also found in molecules,\citep{Dunlap1983_JCP_4997,Dunlap1984_PRA_2902,Schipper1998_TCA_329}
and are likely the main challenge in large-scale black-box applications
of DFT. A reusable implementation of such an approach in OpenOrbitalOptimizer
would then allow exploring fractional-occupation solutions in any
context, given the flexibility of the present approach.

One other potential avenue for future development are interfaces to
other languages, such as Python and Fortran, which is easiest to accomplish
via a C language interface.\citep{Lehtola2023_JCP_180901} We hope
to be able to accommodate a C language interface in the future. While
the C++ library is header-only, the C interface will result in a compiled
binary, which may complicate software distribution.

\section*{Appendix: Generalization of EDIIS to many kinds of particles}

Let us consider the extension of EDIIS\citep{Kudin2002_JCP_8255}
to the case where the energy depends on density matrices $\{\boldsymbol{P}_{\sigma}\}$
of several types of particles $\sigma$, which may be represented
using different basis sets. The Hartree--Fock energy functional,
which is used to build the EDIIS proxy energy, can be summarized as
\begin{align}
E^{\text{HF}}(\{\boldsymbol{P}_{\sigma}\}) & =\sum_{\sigma}\text{Tr }\boldsymbol{h}_{\sigma}\boldsymbol{P}_{\sigma}+\frac{1}{2}\sum_{\sigma}\text{Tr }\boldsymbol{G}_{\sigma}(\boldsymbol{P}_{\sigma})\boldsymbol{P}_{\sigma}\nonumber \\
 & +\frac{1}{2}\sum_{\sigma\neq\sigma'}\text{Tr }\boldsymbol{J}_{\sigma'}(\boldsymbol{P}_{\sigma})\boldsymbol{P}_{\sigma'}\label{eq:uhf}
\end{align}
where $\boldsymbol{h}_{\sigma}$ is the one-electron Hamiltonian,
$\boldsymbol{G}_{\sigma}(\boldsymbol{P}_{\sigma})$ contains the Coulomb
and exchange contributions, and $\boldsymbol{J}$ is just the Coulomb
interaction, which satisfies $\boldsymbol{J}_{\sigma'}(\boldsymbol{P}_{\sigma})\boldsymbol{P}_{\sigma'}=\boldsymbol{J}_{\sigma}(\boldsymbol{P}_{\sigma'})\boldsymbol{P}_{\sigma}$
but we include both terms to make the end result symmetric by construction.
The Fock matrix corresponding to \ref{eq:uhf} is
\begin{equation}
\boldsymbol{F}_{\sigma}=\partial E^{\text{HF}}/\partial\boldsymbol{P}_{\sigma}=\boldsymbol{h}_{\sigma}+\boldsymbol{G}_{\sigma}(\boldsymbol{P}_{\sigma})+\boldsymbol{J}_{\sigma}(\boldsymbol{P}_{\sigma'})\label{eq:fock-uhf}
\end{equation}
and thereby the difference between the Fock matrices of iterations
$j$ and $i$ is
\begin{align}
\boldsymbol{F}_{\sigma}^{(j)}-\boldsymbol{F}_{\sigma}^{(i)} & =\boldsymbol{G}_{\sigma}\left(\boldsymbol{P}_{\sigma}^{(j)}\right)-\boldsymbol{G}_{\sigma}\left(\boldsymbol{P}_{\sigma}^{(i)}\right)\nonumber \\
 & +\boldsymbol{J}_{\sigma}\left(\boldsymbol{P}_{\sigma'}^{(j)}\right)-\boldsymbol{J}_{\sigma}\left(\boldsymbol{P}_{\sigma'}^{(i)}\right).\label{eq:uhf-fock-diff}
\end{align}

In the proof, one merely extends the definition of the EDIIS energy
functional to the case of different types of particles, while still
employing the same extrapolation coefficients $c_{i}\ge0$ with $1=\sum_{i}c_{i}$:
\begin{equation}
f^{\text{EDIIS}}(\boldsymbol{c})=E^{\text{HF}}\left(\left\{ \sum_{i}c_{i}\boldsymbol{P}_{\sigma}^{(i)}\right\} \right).\label{eq:uEDIIS}
\end{equation}
As noted by \citet{Kudin2002_JCP_8255}, the manipulation is straightforward;
it suffices to note that one first needs to insert $1=\sum_{j}c_{j}$
to arrive at an expression which looks like $\sum_{ij}A_{ij}P_{i}$,
where $\boldsymbol{A}$ is an antisymmetric quantity, $A_{ij}=-A_{ji}$.
In the second step, splitting the sum in two copies and interchanging
the indices $i\leftrightarrow j$ in the other term, one gets $\sum_{ij}A_{ij}P_{i}=\frac{1}{2}\sum_{ij}A_{ij}P_{i}+\frac{1}{2}\sum_{ji}A_{ji}P_{j}=\frac{1}{2}\sum_{ij}A_{ij}(P_{i}-P_{j})$.
These two steps are sufficient to manipulate \ref{eq:uEDIIS} into
the form
\begin{align}
f^{\text{EDIIS}}(\boldsymbol{c}) & =\sum_{i}c_{i}E^{\text{UHF}}(\{\boldsymbol{P}_{\sigma}^{(i)}\})\nonumber \\
 & -\frac{1}{4}\sum_{ij}c_{i}c_{j}\sum_{\sigma}\text{Tr }\nonumber \\
 & \left[\boldsymbol{F}_{\sigma}\left(\boldsymbol{P}_{\sigma}^{(j)}\right)-\boldsymbol{F}_{\sigma}\left(\boldsymbol{P}_{\sigma}^{(i)}\right)\right]\times\nonumber \\
 & \left(\boldsymbol{P}_{\sigma}^{(j)}-\boldsymbol{P}_{\sigma}^{(i)}\right)\label{eq:uEDIIS-simple}
\end{align}
which remarkably has the same form as \ref{eq:uEDIIS}, where now
the quadratic term just has a sum over the particle types. This also
proves that EDIIS works more generally, for example with the nuclear-electronic
orbital method, as it suffices to sum over the contributions over
the various particle types. As EDIIS and ADIIS coincide in the case
of Hartree--Fock, the same observation should also hold for ADIIS.

\section*{CRediT author statement}

\textbf{Susi Lehtola}: Conceptualization, Data Curation, Formal Analysis,
Funding Acquisition, Investigation, Methodology, Project Administration,
Resources, Software (lead), Supervision, Validation, Writing -- Original
Draft Preparation, Writing -- Review \& Editing (lead).

\textbf{Lori Burns}: Software (supporting), Writing -- Review \&
Editing (supporting).

\section*{Acknowledgments}

We dedicate this work to the memory of John F Stanton, who was a positive
influence on the community and a supporter of connecting software.
SL thanks Eric Cancès for many discussions on orbital optimization
and the optimal damping algorithm, Mi-Song Dupuy for discussions on
the DIIS algorithm, Antoine Levitt for discussions on finite precision
issues, his brother Ville Lehtola for discussions on reproducible
science, as well as Alejandro Garza and Gustavo Scuseria for confirming
the form of the EDIIS+DIIS interpolation function. SL thanks the Academy
of Finland for financial support under project numbers 350282 and
353749. SL also acknowledges the critical support of a French--Finnish
Maupertuis researcher short mobility grant funded by the French Institute
in Finland, the French Ministry of Higher Education and Research,
and the Finnish Society of Sciences and Letters at an early stage
of this project.

\begin{tocentry}
\includegraphics{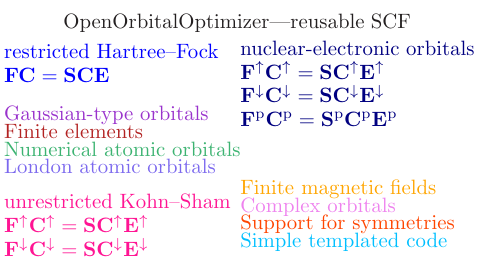}
\end{tocentry}
\clearpage{}

\bibliography{citations}

\end{document}